\documentclass[twocolumn,nofootinbib]{revtex4-1}
\usepackage{latexsym,epsfig,amssymb, amsmath,nicefrac}

\usepackage{amsfonts,amsthm} 
\usepackage{bm,bbm}
\usepackage{graphicx}
\usepackage{esint}
\usepackage{color}
\usepackage{braket}

\newcommand{\be}{\begin{equation}}
\newcommand{\ee}{\end{equation}}
\newcommand{\ba}{\begin{eqnarray}}
\newcommand{\ea}{\end{eqnarray}}
\def\bs{\begin{subequations}}
\def\es{\end{subequations}}

\def\vp{\varphi}

\def\Kahler{K\"{a}hler~}
\def\Mp{M_{\rm Pl}}

\newcommand{\rf}[1]{(\ref{#1})}

\begin{document}

\title{{\Large A universal attractor for inflation at strong coupling}}

\author{Renata Kallosh${}^1$, Andrei Linde${}^1$ and Diederik Roest${}^2$}

\affiliation{{}$^1$Department of Physics and SITP, Stanford University, \\ 
Stanford, California 94305 USA, kallosh@stanford.edu, alinde@stanford.edu}

\affiliation{{}$^2$Centre for Theoretical Physics, University of Groningen, \\ Nijenborgh 4, 9747 AG Groningen, The Netherlands, d.roest@rug.nl}

\begin{abstract}
We introduce a novel non-minimal coupling between gravity and the inflaton sector. Remarkably, for large values of this coupling all models asymptote to a universal attractor. This behavior is independent of the original scalar potential and generalizes the attractor in the $\phi^4$ theory with non-minimal coupling to gravity.  The attractor is  located in the `sweet spot' of Planck's recent results.
\end{abstract}

\maketitle

\smallskip

\noindent {\bf Introduction.\ } 
The data releases by WMAP9 and Planck2013 \cite{Hinshaw:2012aka} attracted attention of cosmologists to two very different cosmological models which, surprisingly, made very similar observational predictions: the Starobinsky model $R+R^{2}$ \cite{Starobinsky:1980te} and the chaotic inflation model $V(\phi) \sim \phi^{4}$ \cite{Linde:1983gd} with non-minimal coupling to gravity ${\xi\over 2}\phi^{2} R$  \cite{Salopek:1988qh,Bezrukov:2013}. For $\xi \gtrsim 0.1$, both of these models predict that for large number of e-foldings $N$, the spectral index and tensor-to-scalar ratio are given by
 \be  n_s  = 1-2/N \,, \quad 
  r =12/N^2 \, . \label{observables}
\ee
For $N \sim 60$, these predictions $n_s   \sim 0.967$, $r  \sim 0.003$ ($n_s   \sim 0.964$, $r  \sim 0.004$ for $N \sim 55$) are in the sweet spot of the WMAP9 and Planck2013 data.

Further investigations revealed that many other inflationary theories also predict $n_{s} $ and $r$ given by (\ref{observables}). In particular, (\ref{observables}) is a universal attractor point for a broad class of theories with spontaneously broken conformal or superconformal invariance \cite{Kallosh:2013hoa}, and for closely related models with negative non-minimal coupling $\xi < 0$ \cite{Kallosh:2013maa}. 
 
 However, until now, in the theories with non-minimal coupling ${\xi\over 2}\phi^{2} R$ with $\xi > 0$, this generality did not  extend beyond the models with the potentials $\sim \phi^{4}$  studied in \cite{Salopek:1988qh,Bezrukov:2013}.  
In this paper, we propose a very simple generalization of this class of models, which applies to practically every inflationary potential $V(\phi)$. This can be achieved by introducing a generalized version of non-minimal coupling to gravity, such as  $\xi \sqrt{V(\phi)} R$, or even a much simpler one, $\xi\phi R$. We will show that all of these models have the universal set of predictions (\ref{observables})  in the strong coupling limit  $\xi\to \infty$. We will also show how exactly the predictions of the theories with different potentials $V(\phi)$ depend on $\xi$ and approach the universal attractor point (\ref{observables}) with the growth of $\xi$.

\smallskip

\noindent {\bf Non-minimal coupling.\ } 
The starting point of many inflationary models is a Lagrangian consisting of the Einstein-Hilbert term for gravity plus a kinetic term and scalar potential for the inflaton field. The Lagrangian including the generalized non-minimal coupling to gravity reads
  \begin{align}
 & \mathcal{L}_{\rm J} = \sqrt{-g} [ \tfrac12  \Omega(\phi) R  - \tfrac12 (\partial \phi)^2 -  V_J(\phi) ] \,, \label{Baction} 
\end{align}
with\footnote{Various aspects of generalized non-minimal coupling were studied in \cite{Barbon:2009ya,Pallis,Qiu,Chakravarty,Kaiser}.}
\begin{align}
& \Omega(\phi)  = 1 + \xi f(\phi)\, ,  \qquad V_J(\phi)=\lambda^2 f^2(\phi) \,.
\end{align}
Our notation for $V_J(\phi) $ does not imply any constraint on the scalar potential other than being positive, and is motivated by the superconformal version of the model that will be introduced later. Due to the non-minimal coupling, we will refer to this form of the theory as Jordan frame. In order to transform to the canonical Einstein frame, one needs to redefine the metric:
 \begin{align}
  g_{\mu \nu} \rightarrow \Omega(\phi)^{-1} g_{\mu \nu}  \,.
 \end{align}
This bring the Lagrangian to the Einstein-frame form:
 \begin{align}
 & \mathcal{L}_{\rm E} = \sqrt{-g} [    \tfrac12   R - \tfrac12 \Big (\Omega(\phi)^{-1} + \tfrac32 (\log \Omega(\phi))'^2\Big ) (\partial \phi)^2 + \notag \\
& - V_E(\phi)] \,, \qquad {\rm with~~} V_E(\phi) =  \frac{V_J(\phi)}{\Omega(\phi)^2}  \,. \label{Einstein}
 \end{align}
Note that in the absence of non-minimal coupling, $\xi = 0$, the distinction between Einstein and Jordan frame vanishes. In this case the inflationary dynamics is fully determined by the properties of the scalar potential $V_J(\phi)= V_E(\phi)$. In the presence of a non-minimal coupling, however, one has to analyze the interplay between the different contributions  to the inflationary dynamics due to $V_J(\phi)$ and $\xi$. 

\smallskip

\noindent {\bf Behavior at weak coupling.\ }
We first analyze the effect of the non-minimal coupling for small $\xi$. At linear order, the kinetic terms in \eqref{Einstein} give rise to the following definition of the canonical scalar field $\varphi$:
 \begin{align}
 \frac{\partial \varphi}{\partial \phi} = 1 - \tfrac\xi 2 f(\phi) \,,
 \end{align}
where we are suppressing higher-order terms. A similar approximation can be made to the Einstein-frame potential,
 \begin{align}
  V_E = \lambda^2f(\phi)^2 (1 - 2 \xi f(\phi)) \,.
 \end{align}
Remarkably, this implies that the number of e-foldings 
 \begin{align}
  N = \int^{\phi_N}_{\phi_{\rm end}} \left(  \frac{\partial \varphi}{\partial \phi} \right)^2 \frac{V_E}{\partial V_E / \partial \phi} d \phi \,, 
 \end{align}
has no linear corrections. There will be corrections to $N$ due to changes to the field value $\phi_{\rm end}$ when inflation breaks down since $\epsilon$ or $\eta$ become of order one. However, these will be subdominant as $N$ generically receives the largest contribution from the first phase of the inflationary trajectory, where $\partial V / \partial \phi$ is small. At first approximation, there are therefore no changes to $N$ at linear order. The only corrections to the slow-roll parameters follow from the explicit expressions for these quantities,
 \begin{align}
   \epsilon &  = \frac{1}{2}  \left(   \frac{1}{V_E} \frac{\partial V_E}{\partial \phi} \frac{\partial \phi}{\partial \varphi} \right)^2 = (1  - \xi f(\phi_N)) \epsilon_J \,, \notag \\
  \eta & =   \frac{1}{V_E} \frac{\partial}{\partial \phi}  \left(  \frac{\partial V_E}{\partial \phi} \frac{\partial \phi}{\partial \varphi} \right)  \frac{\partial \phi}{\partial \varphi} = \eta_ J - \tfrac52 \xi f(\phi_N) \epsilon_J \,, 
 \end{align}
evaluated at the same point in field space $\phi_{N}$ as for the original scalar potential $V_J(\phi)$. Given a value $(r_0, n_{s0})$ for the cosmological observables of any inflationary model without non-minimal coupling, at small coupling these will transform in the following universal way:
 \begin{align}
  n_{s} & = 1 +2 \eta - 6 \epsilon = n_{sJ} + \tfrac{\xi}{16} f(\phi) r_{J} \,, \notag \\
r &= 16 \epsilon = r_J - \xi f(\phi) r_J \,. \label{weak-coupling}
 \end{align}
Therefore all models at first will move along parallel lines with a slope of $-16$ in the $(n_s,r)$-plane.

\smallskip

\noindent {\bf Behavior at strong coupling.\ }
Next we turn to the strong coupling limit of inflation, where $\xi$ becomes very large. We will later quantify how large $\xi$ needs to be for this limit. First we will present two arguments for a universal attractor behavior in the limit of infinite $\xi$. The first argument follows the line of reasoning above, but considers an expansion for large $\xi$ instead. The number of e-foldings in this case reads
 \begin{align}
 N =  \int^{\phi_N}_{\phi_{\rm end}} \left( \frac{3}{4} \xi f(\phi)' + \frac{f(\phi)}{2 f(\phi)'}- \frac{3 f(\phi)'}{4 f(\phi)} \right) d \phi \,. \label{N-strong-coupling}
 \end{align}
Without specifying the function $f(\phi)$, the first term can be integrated in a model-independent way; this would not be possible when including next-to-linear order terms. Here we assume that we are away from the extrema of $f$ where $f'=0$ so that the second term in \rf{N-strong-coupling}
 blows up. Moreover, one can neglect the contribution from the end of inflation in the large-$N$ limit (this is also true at strong coupling). We therefore obtain 
\be\label{N}
N = \tfrac34 \xi f(\phi_N) \,.
\ee
Note that this requires $f(\phi_N)$ to asympote to zero in the strong coupling limit; one zooms in on the region where the scalar potential vanishes. In this limit one obtains the simple formula for the spectral index and tensor-to-scalar ratio (\ref{observables}).
This analysis demonstrates that the values of $n_{s}$ and $r$ for all positive scalar potentials $V_J(\phi)$ with a Minkowski minimum asymptote to (\ref{observables}) in the strong coupling limit. 

The second argument starts from the kinetic term in Einstein frame \eqref{Einstein}. In the large-$\xi$ limit, the two contributions to the kinetic terms scale differently under $\xi$. Retaining only the leading term, the Lagrangian becomes
  \begin{align}
  \mathcal{L}_{\rm E} = \sqrt{-g} \bigg[ & \tfrac12  R - \tfrac34 (\partial \log (\Omega(\phi)))^2 
  - \lambda^2 \frac{f(\phi)^2}{\Omega(\phi)^2}  \bigg] .
 \end{align}
Remarkably, the canonically normalized field $\varphi$ involves the function $\Omega(\phi)$ of the scalar potential itself:
 \begin{align}
  \varphi = \pm \sqrt{\nicefrac32} \log (\Omega(\phi)) \,. \label{canonical}
 \end{align}
Therefore, in terms of $\varphi$, the theory has lost all reference to the original scalar potential, it has the universal form. In case of odd $f(\phi)$ we choose the same sign in \eqref{canonical} for both signs   of $\varphi$ and find
  \begin{align}\label{potstar}
  \mathcal{L}_{\rm E} = \sqrt{-g} \left[ \tfrac12   R - \tfrac12 (\partial \varphi)^2 - \frac{ \lambda^2}{\xi^{2}} (1 - e^{-\sqrt{{2\over 3}} \, \varphi})^{2}  \right] \,,
 \end{align}
which is the scalar formulation of the Starobinsky model \cite{Starobinsky:1980te}. 
In case the function $f(\phi)$ is  even in $\phi$ we choose opposite signs and find the following attractor action
  \begin{align}\label{pothiggs}
  \mathcal{L}_{\rm E} = \sqrt{-g} \left[ \tfrac12   R - \tfrac12 (\partial \varphi)^2 - \frac{ \lambda^2}{\xi^{2}} \Big (1 - e^{-\sqrt{{2\over 3}  \varphi^2}} \, \, \Big )^{2}  \right] ,
 \end{align}
symmetric under $\vp \rightarrow -\vp$. 

The crucial assumption in the above derivation was that the kinetic term is dominated by the second contribution. In other words, we require
 \begin{align}
  \Omega(\phi) \ll \tfrac32 \Omega(\phi)'^2 \,. \label{approx}
 \end{align}
In terms of our original scalar potential and the associated slow-roll parameter $\epsilon_{\rm J}$, this translates into 
 \begin{align}
  \frac{1 + \xi f(\phi)}{\xi^2 f(\phi)^2} \ll \tfrac34  \epsilon_{\rm J} (\phi) \,.
 \end{align}
Interestingly, this implies that models with a flatter scalar potential require a stronger coupling in order to reach the vicinity of the attractor. In contrast, for less fine-tuned models with larger values of $\epsilon_{\rm J}$, the system reaches the attractor for a lower value of the coupling $\xi$. It is important to point out that even models with a scalar potential that does not support inflation,  still asymptote to \rf{potstar} or  \rf{pothiggs} at strong coupling and have the same observables \rf{observables}.

The amplitude normalization of the power spectrum constrains the overall coefficient of the scalar potentials. For $\xi = 0$ this depends on the coefficient $\lambda$ of the original scalar potential. For large $\xi$, the Planck normalization of the power spectrum requires $\lambda/\xi \approx 10^{-5}$. For intermediate values there is an interplay between the coefficients $\lambda$ and $\xi$, which can always be satisfied by suitable choice of  $\lambda$. For the specific case of the $\phi^4$ theory this was discussed in detail in \cite{Bezrukov:2013}.

\smallskip

\noindent {\bf Supergravity embedding.\ }
The non-minimal coupling can be embedded in supergravity. We follow the set-up of \cite{Rube:2010}, which introduces two chiral multiplets with scalar fields $\Phi$ and $S$. The former will contain the inflaton while the latter is responsible for SUSY breaking. We thus take the sGoldstini to be orthogonal to the inflaton, allowing for an arbitrary scalar potential and avoiding the restrictions of \cite{Achucarro}. While the original proposal has a specific \Kahler potential and an arbitrary function in the superpotential, we take the \Kahler potential to depend on $\Omega(\sqrt{2} \Phi)$ which will be related to the scalar potential. Our final expressions are:
 \begin{align}
\hskip -10pt K =  & - 3 \log[ \tfrac12 ( \Omega(\sqrt{2} \Phi) + \Omega(\sqrt{2} \bar \Phi))   - \tfrac13 S \bar S + \tfrac16 (\Phi - \bar \Phi)^2 \notag \\
& +  \zeta \frac{( S \bar S)^2}{\Omega(\sqrt{2} \Phi) + \Omega(\sqrt{2} \bar \Phi)}] \,, \quad
  W = \lambda S f(\sqrt{2} \Phi)  \,,
\label{KW} \end{align}
where $\Omega(\sqrt{2} \Phi) = 1 + \xi f(\sqrt{2} \Phi)$ and $f(\sqrt{2} \Phi)$ is a real holomorphic function. This leads exactly to the bosonic model discussed above upon identifying $\Phi = \phi / \sqrt{2}$ while $S=0$. 
It can easily be seen that this is a consistent truncation. 

The superconformal version of this model explains the simplicity of the Jordan frame potential in these models:  in a gauge where the conformon is fixed, the superconformal potential is given by ${\cal W} =  \lambda S  f  ({\sqrt 2 }\, \Phi )$ (in the notation of \cite{Ferrara:2010in, Kallosh:2013pby}). This implies that the Jordan frame potential at  $S=0, \Phi= \phi/{\sqrt 2}$, is  given by
\be
V_J= \lambda^2 \Big |{\partial {\cal W}\over \partial  S}\Big |^2 = \lambda^2 f^2(\phi) \,.
\ee This model generalizes the supersymmetric embedding of the $\phi^4$ theory considered in \cite{Kallosh:2013pby} to arbitrary scalar potentials. In that specific case, one could interpolate between a canonical \Kahler potential depending on $\Phi \bar \Phi$ and a shift-symmetric one depending on $(\Phi - \bar \Phi)^2$ by means of $\xi$, but this is not possible in the general case.

Regarding the stability of the truncation to the inflationary trajectory, where three scalars are truncated out, the masses of the four fields are given by
$ m_{{\rm Re}\, \Phi}^2  = \eta V$, 
  $m_{{\rm Im}\, \Phi}^2  = ( \nicefrac43 + 2 \epsilon - \eta ) V$,  $m_S^2  = ( - \nicefrac23 + 6 \zeta + \epsilon ) V$.
Up to slow-roll corrections, one can thus stabilize all three truncated fields with the choice $\zeta > \nicefrac{1}{9}$.

This supergravity embedding goes some way towards an understanding of the symmetries underlying the attractor behavior. In particular, for $\xi = 0$ there is symmetry enhancement in the \Kahler potential: it has a shift symmetry in the real part of $\Phi$ and hence does not depend on the inflaton. The same holds for any value of $\xi$ when choosing the function $f(\sqrt{2} \Phi)$ to be a constant. Any deviations from this will introduce a spontaneous breaking of this symmetry.

\begin{figure}[t!]
\vspace{-.3cm}
\centerline{\includegraphics[scale=.4]{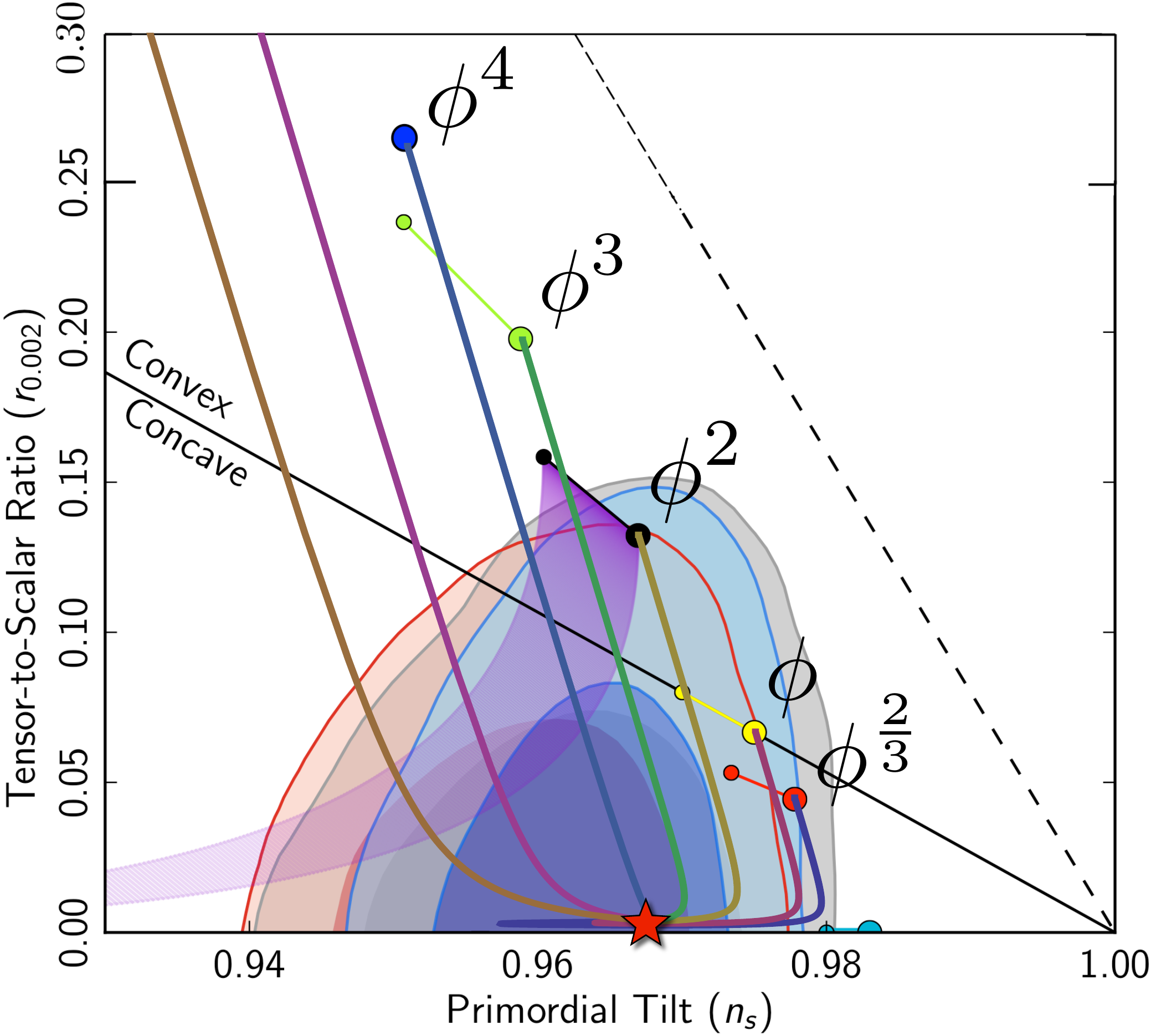}}
\centerline{\includegraphics[scale=0.42]{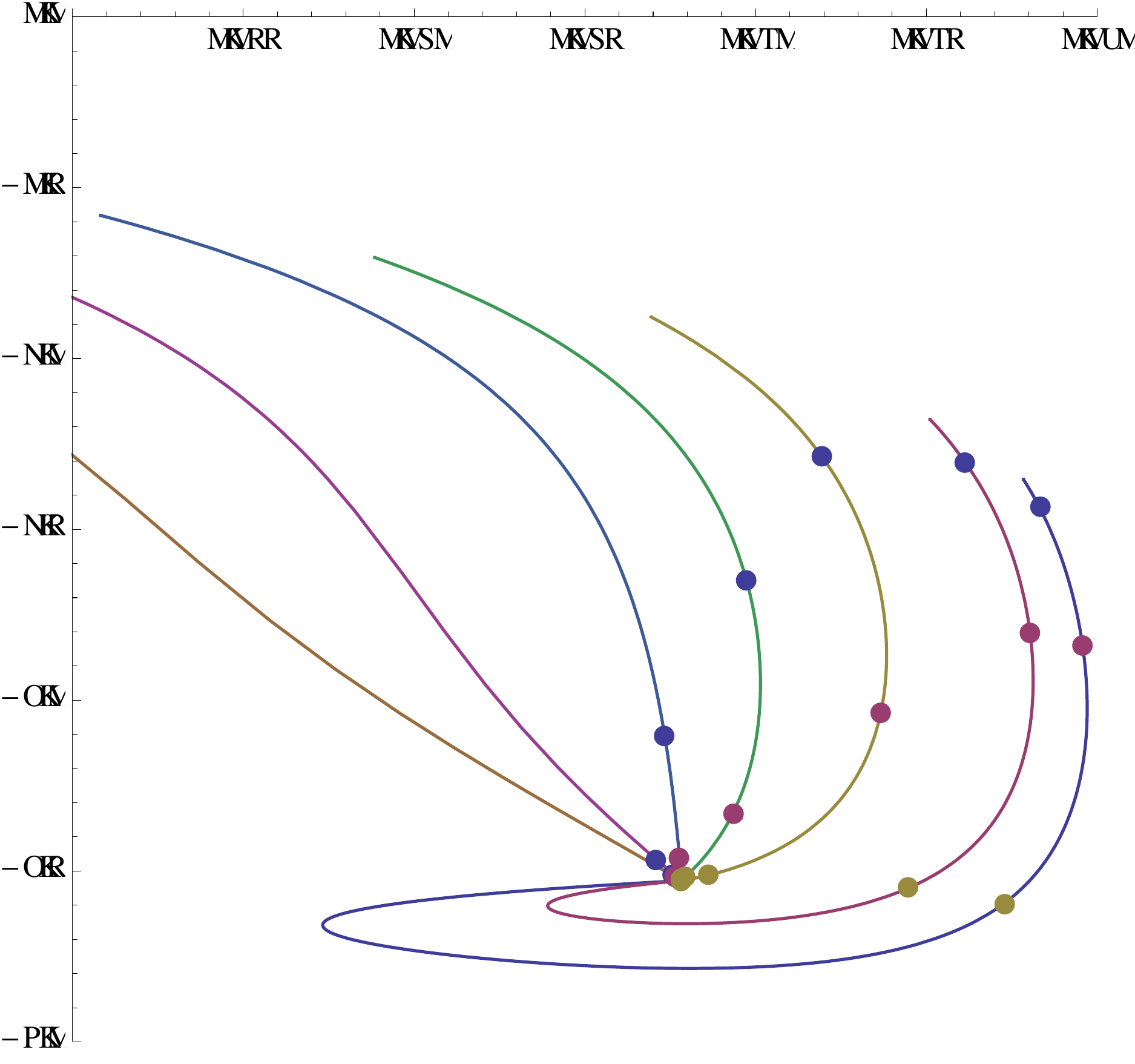}}
\caption{\it \small{The $\xi$-dependence of $(n_s,r)$ on a linear and a logarithmic scale for different chaotic models with $n=(2/3, 1, 2, 3, 4, 6, 8)$, from right to left, for 60 e-foldings. The points on the logarithmic scale (lower panel) correspond to $\log(\xi)=(-1, \ldots, 1)$, from top down. The convergence to the attractor point occurs almost instantly for $n \geq 4$.}}
\vspace{-.3cm}
\end{figure} 

\smallskip

\noindent {\bf Chaotic inflation.\ }
In this section we illustrate the universal attractor behavior for chaotic inflation \cite{Linde:1983gd}, with the scalar potential  \begin{align}
  V_J(\phi) =  \lambda^2 \Mp^{4-n} \phi^n \,.
 \end{align}
Without non-minimal couplings, these have the following cosmological observables:
 \begin{align}
  n_{sJ} = 1 - \frac{2+n}{2N} \,, \qquad
  r_J = \frac{4n}{N} \,,
 \end{align}
at large $N$. These are specific cases of the most general $1/N$-dependence derived in  \cite{Mukhanov:2013}. The attractor behavior for this class is depicted in Figure 1. The cross-over behavior between the two regimes spans a number of decades of the non-minimal coupling $\xi$, and in addition is model-dependent. Indeed models with a larger $\epsilon_{\rm J}(\phi)$ require a smaller coupling to approach the attractor. However, the attractor is always reached before $\xi = 100$.  Importantly, for $n = 4$ the attractor is reached very early, for $\xi \gtrsim 10^{-1}$, and for $n > 4$ the convergence occurs even much faster, for much smaller values of $\xi$, see Figure 1. In this sense, the words ``strong coupling limit'' are not entirely adequate, because the ``strong coupling regime'' occurs in these models very early, at $\xi \ll 1$.

\smallskip
\noindent {\bf Unitarity bound?\ }
The models discussed above significantly generalize the chaotic inflation model $\lambda\phi^{4}/4$ with non-minimal coupling to gravity ${\xi\over 2}\phi^{2} R$, which was used in \cite{Salopek:1988qh} for the discussion of the Higgs inflation scenario with $\lambda = O(1)$, $\xi \sim 10^{5}$. In this respect we should mention that during the discussion of the Higgs inflation models, several authors claimed that Higgs inflation suffers from the unitarity problem \cite{Burgess:2009ea,Barbon:2009ya,Hertzberg:2010dc,Germani:2010gm}, whereas some others argued that this problem does not exist  \cite{Ferrara:2010in,Bezrukov:2010jz}. 

Here we will briefly discuss this issue and check whether a similar issue arises in the new class of models as well \cite{Kehagias:2013mya,Giudice:2014toa}. 
A more precise statement of the result obtained in \cite{Burgess:2009ea,Barbon:2009ya,Hertzberg:2010dc,Germani:2010gm}  is that in the vicinity of the Higgs minimum, at $\vp \ll 1/\xi \sim O(10^{-5})$, higher order quantum corrections to scattering amplitudes become greater than the lower order effects for energies greater than $O(1/\xi)$. The energy scale $O(1/\xi)$ above which the perturbation theory fails at $\vp \ll 1/\xi$ was called the unitarity bound. Since the Hubble constant during inflation is $O(\sqrt\lambda/\xi)$, and $\lambda = O(1)$ in Higgs inflation, it was conjectured in  \cite{Burgess:2009ea,Barbon:2009ya,Hertzberg:2010dc,Germani:2010gm} that the description of the Higgs inflation using perturbation theory is unreliable. However, inflation happens at $\vp \geq O(1)$, which is $10^{5}$ times greater than the range of the values of the field where the existence of the problem was established. An investigation performed in \cite{Ferrara:2010in,Bezrukov:2010jz} demonstrated that the higher order corrections are negligible during inflation because at  large $\vp$ the potential \rf{pothiggs} is exponentially flat and the effective coupling constant $\lambda(\vp)$ is exponentially small.

After inflation, when the field $\vp$ becomes very small, one may encounter problems in describing reheating by perturbation theory. But is it a real problem? It is well known that the perturbative approach to reheating fails in many inflationary models anyway, which requires using non-perturbative methods developed in  \cite{KLS,latticeold}. This does not affect inflationary predictions and does not lead to any problems with the inflationary scenario. 

This does not mean that the unitarity issue is entirely inconsequential.  The problem may re-appear if one attempts to develop LHC-related particle phenomenology models with the Higgs field playing the role of the  inflaton. This may require solving RG equations up to the Planckian energies, which is problematic for $\xi \gg 1$. But this is not a problem of consistency of inflationary models but rather a specific problem of particle phenomenology beyond the Standard Model. An interesting way to avoid this problem was recently proposed in \cite{Giudice:2014toa}. However, the mass of the inflaton field in the class of the universal attractor models developed in  \cite{Giudice:2014toa} is 10 orders of magnitude greater the Higgs mass, so this approach is not directly related to Higgs inflation and particle phenomenology beyond the Standard Model.

In this paper, we do not make any attempts to relate the inflaton field to the Higgs field discovered at LHC. In particular, the coupling constant $\lambda$ in the generic chaotic inflation models $\lambda\phi^{4}/4$ with non-minimal coupling to gravity ${\xi\over 2}\phi^{2} R$ can be extremely small. This makes the inflationary energy scale $H \sim \sqrt\lambda/\xi$ much smaller than $O(1/\xi)$ and alleviates the problems discussed in \cite{Burgess:2009ea,Barbon:2009ya,Hertzberg:2010dc,Germani:2010gm}. Moreover, in the theories $\phi^{n}$ with $n\lesssim 1$ the unitarity bound is much higher than the Planck mass even for large $\xi$ \cite{Kehagias:2013mya}. For $n\geq 2$, our results show that the attractor behavior occurs starting from $\xi <1$, in which case the unitarity bound is also super-planckian. Therefore we believe that the perturbative unitarity problem does not affect the main results obtained above.

\smallskip

\noindent {\bf Natural inflation.\ }
As a second example, we consider natural inflation \cite{natural} with
 \begin{align}
  V_J(\phi) = \lambda^2 \Mp^4 (1 + \cos(\phi/(f \Mp)))\,.
 \end{align}
This case does not allow for an insightful large-$N$ expansion due to the presence of non-perturbative terms. The perturbative answer is independent of $N$:
 \begin{align}
  n_{sJ} = 1 - \frac{1}{f^2} \,, \quad r_J = 0 \,,
 \end{align}
and hence falls outside of the scope of \cite{Mukhanov:2013}. Again, irrespective of the value of $f$, this model asymptotes to the same attractor, see figure 2. Also in this case the cross-over takes a number of decades of $\xi$, but is always completed by the time $\xi$ reaches $1000$.

\begin{figure}[t!]
\vspace{-.3cm}
\centerline{\includegraphics[scale=0.55]{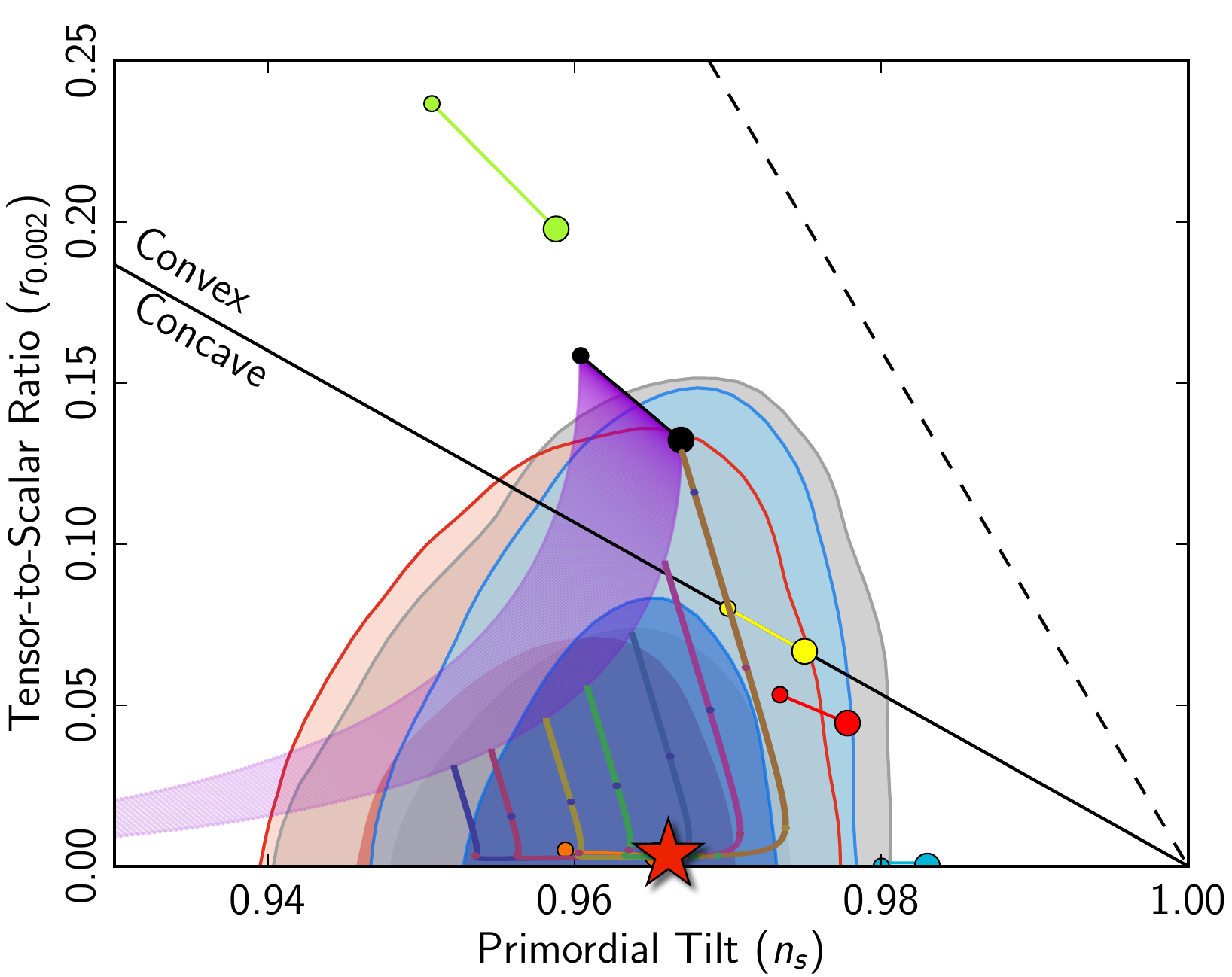}}
\centerline{\includegraphics[scale=1]{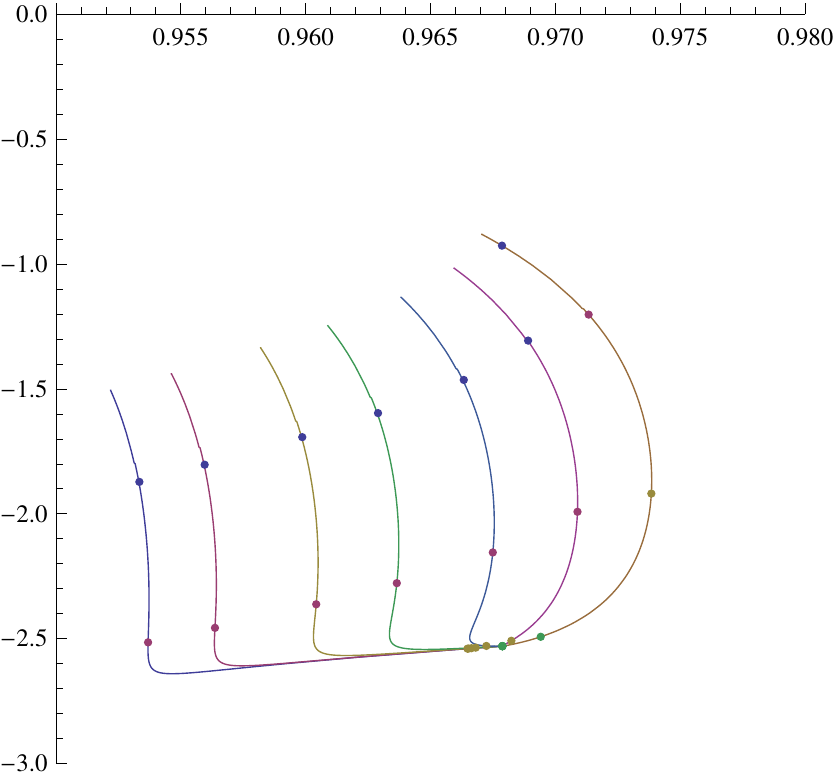}}
\caption{\it \small{The $\xi$-dependence of $(n_s,r)$ on a linear and a logarithmic scale for different natural models with $f=(5, 5.25, 5.75, 6.33, 7.5, 10, 100)$ (in decreasing redshift) for 60 e-folds. The points correspond to $\log(\xi)=(0,\ldots,3)$.}}
\vspace{-.3cm}
\end{figure}

\smallskip

\noindent {\bf Generalized strong coupling attractor.\ }
In the previous investigation, we assumed that $\Omega(\phi) = 1 + \xi f(\phi)$, and  $V_J(\phi)=\lambda^2 f^2(\phi)$, but one may also consider a more general possibility,
\be
\Omega(\phi) = 1 + \xi g(\phi)\, ,  \qquad V_J(\phi)=\lambda^2 f^2(\phi) \, , 
\ee
where we introduce an additional functional freedom in the definition of $\Omega(\phi)$, disconnecting it from $V_J(\phi)$. Once we do so, $V_E(\phi) =  V_J(\phi) / \Omega^{2}(\phi)$ no longer approaches a constant at large $\phi$. Does it mean that our previous results become inapplicable?

Note that when the field rolls to the minimum of its potential, $f(\phi)$ is supposed to vanish, or at least to become incredibly small to account for the incredible smallness of the cosmological constant $\sim 10^{-120}$. As in the previous analysis, we will assume that the same is true for the function $g(\phi)$. Therefore we will expand both functions in a Taylor series in $\phi$, assuming that they vanish at some point (which can be taken as $\phi =0$ by a field redefinition) and that they are differentiable at this point:
 \be
f(\phi) = \sum_{n=1}^{\infty} f_{n} \phi^{n} \ , \qquad g(\phi) = \sum_{n=1}^{\infty}  g_{n} \phi^{n} \ . 
 \ee
By rescaling $\lambda $ and $\xi$, one can always redefine $f_{1} = g_{1} = 1$ without loss of generality.

Let us first ignore all higher order corrections, i.e. take $f(\phi) = g(\phi) = \phi$. In this case our investigation is reduced to the one performed earlier, and equation (\ref{N}) yields $\phi_{\rm N} = {4N\over 3\xi}$. This result implies that for $\xi \gg N$ one has $\phi_{\rm N}\ll 1$.

If one now adds all higher order terms and makes an assumption that  the coefficients $f_{n}$ and $g_{n}$ are $O(1)$, one finds that in the large coupling limit $\xi \gg N$, these corrections are suppressed by the powers of ${4N\over 3\xi}$, so one can indeed ignore these terms. This means that in the large $\xi$ limit the potential $V(\varphi)$ in terms of the canonically normalized inflaton field $\varphi$ coincides with the potential (\ref{potstar}), and  all observational predictions of this new class of theories coincide with the predictions (\ref{observables}). This universality is similar to the universality of predictions of the broad class of theories found in \cite{Kallosh:2013hoa,Kallosh:2013maa}.

In this analysis we assumed that the Taylor series begins with the linear term. However, if the theory is symmetric with respect to the change $\phi \to -\phi$, as is the case e.g.~in the $\phi^4$ theory, then the expansion for $f(\phi)$ and $g(\phi)$ begins with the quadratic terms. The rest follows just as in the case discussed above: For  $\xi\gg N$, higher order corrections do not affect the description of the observable part of the universe, we have the same observational predictions (\ref{observables}) as before, but now the relevant part of the potential is even with respect to the field $\varphi$ and its large $\xi$ limit is given by (\ref{pothiggs}).
  
\smallskip

\noindent {\bf Discussion.\ }
In this letter we have demonstrated that there is a universal attractor for all inflationary models when introducing a specific non-minimal coupling term correlated with the choice of the potential. Upon taking its coefficient $\xi$ large enough, all models asymptote to a spectral index and tensor-to-scalar ratio that are indistinguishable from  (\ref{observables}), and hence are in perfect agreement with the Planck results. How large $\xi$ needs to be in order to reach the attractor is model-dependent, but in all examples we have found that $\xi = 100$ is sufficient. Moreover, the initial approach to the attractor proceeds in a parallel fashion; upon turning on $\xi$, the different models move in identical directions in the $(n_s,r)$ plane. The resulting image in figure 1 resembles that of a comb. The straight line towards the attractor for the $\phi^4$ theory  is a coincidence between the slope of the lines and the location of that particular theory; other models do not start moving in the direction of the attractor. 

The new class of cosmological attractors (\ref{Baction}) can be generalized in many different ways. We discussed its supergravity/superconformal generalization, as well as the possibility to use the function $\Omega$ not related to $f(\phi)$. This additional universality appears because in the large $\xi$ limit the description of the last $N$ e-foldings of inflation requires knowledge of a very limited range of values of $f(\phi)$ and $\phi$, where the simplest linear or quadratic approximations may be sufficient. Zooming to this limited range of variation of $\phi$ is accompanied by the effective stretching of the potential in terms of the canonical inflaton field $\varphi$. This stretching allows the existence of an inflationary regime even in the theories where the original potential $V_{J}(\phi)$ is very steep. The resulting Einstein frame potential acquires the  form (\ref{potstar}), (\ref{pothiggs}), which leads to the universal observational predictions  (\ref{observables}) for this new class of theories.

\smallskip 
\noindent {\bf Acknowledgements}
We acknowledge stimulating discussions with Eva Silverstein. RK and AL are supported by the SITP and by the NSF Grant PHY-1316699. DR would like to thank the SITP for its warm hospitality and NWO for financial support with a VIDI grant.


\providecommand{\href}[2]{#2}\begingroup\raggedright\endgroup

\end{document}